# Bank financial stability, bank valuation and international oil prices: Evidence from listed Russian public banks

C.T. ALBULESCU[ab*]

[a] *Management Department, Politehnica University of Timisoara, 2, P-ta. Victoriei, 300006, Timisoara, Romania.*

[b] *CRIEF, University of Poitiers, 2, Rue Jean Carbonnier, Bât. A1 (BP 623), 86022, Poitiers, France.*

**Abstract**

Using data on 17 listed public banks from Russia over the period 2008 to 2016, we analyze whether international oil prices affect the bank stability in an oil-dependent country. We posit that a decrease in international oil prices has a negative long-run macroeconomic impact for an oil-exporting country, which further deteriorates the bank financial stability. More specifically, a decrease in international oil prices leads for an oil-exporting country as Russia to a currency depreciation and to a deterioration of the fiscal stance. In addition, given the positive correlation of oil and stock prices documented by numerous previous studies, a decrease in international oil prices represents a negative signal for the stock markets investors, negatively affecting banks' share prices and thus, their capacity to generate sustainable earnings. In this context, the bank financial stability can be menaced. With a focus on public listed banks and using a Pool Mean Group (PMG) estimator, we show that an increase in international oil prices and in the price to book value ratio has a long-run positive effect on Russian public banks stability, and conversely. While positive oil-price shocks contribute to bank stability in the long run, an opposite effect is recorded for negative shocks. However, no significant impact is documented in the short run. Our findings are robust to different bank stability specifications, different samples and control variables.

[*]Corresponding author. E-mail addresses: claudiu.albulescu@upt.ro, claudiual@yahoo.com. Tel: +40 743 089 759.
*Acknowledgements: This work was supported by a grant of the Romanian National Authority for Scientific Research and Innovation, CNCS – UEFISCDI, project number PN-III-P1-1.1-TE-2016-0142.*

# 1. Introduction

The banking sector stability was challenged by the recent financial global crisis all over the world. In the case of Russia, the threats to the bank stability were amplified by the conflict with the Ukraine and the economic sanctions imposed by the Western partners, which leaded to a severe depreciation of the Rubble. Since January 2014, the Russian ruble lost in two years around 50% of its value against the US dollar (Dreger et al., 2016). In addition, the plunge in global oil prices after 2014 put further strain on Russia (Tuzova and Qayum, 2016). In this context, the Russian economy, which is highly responsive to oil price fluctuations, recorded a negative dynamic, the investment and consumption contracted, and the volume of non-performing loans (NPL) increased, putting thus pressure on the banking sector stability.

After 2014, the Russian banking sector was confronted with a considerable growth in credit risks and bad loans, in the context of a severe deterioration of large borrowers' financial condition, and of investors' expectations.[1] The largest share of non-performing loans was recorded in the construction sector, where foreign currency denominated loans prevailed. However, according to the Bank of Russia, the decline in international oil prices generated by an increased oil production at global level, and by the United States monetary policy tightening, put additional pressure on the Russian banking system.[2] The oil and gas industry is of crucial importance for the Russian economy, representing more than 20% of GDP, about 30% of fiscal revenues and more than 50% of exports (Simola and Solanko, 2017). Therefore, the impact of oil prices on the financial stability cannot be neglected. Our purpose is to test the short- and long-run influence of international oil prices on the bank stability in Russia.

The literature addressing the bank stability determinants usually focuses on the role of bank competition (Keeley, 1990; Boyd and De Nicolo, 2005; Martinez-Miera and Repullo, 2010; Fernández et al., 2016), foreign ownership (Lee and Hsieh, 2014), shareholder diversification (García-Kuhnert et al., 2015), non-traditional banking activities (De Jonghe, 2010; Wagner, 2010; Duport et al., 2018; Fina Katmani, 2018) and regulatory framework (Ahamed and Mallick, 2017). A recent strand of the literature (Miyajima, 2016; Khandelwal et al., 2016; Al-Khazali and Mirzaei, 2017; Lee and Lee, 2018) investigate the role of the international oil prices in influencing bank stability.[3] We add to this narrow book of literature by assessing the impact of oil prices on the stability of listed Russian public banks over the period 2008 to 2016.

---

[1] https://www.cbr.ru/Collection/Collection/File/8376/fin-stab-2014-15_4-1_e.pdf
[2] https://www.cbr.ru/Collection/Collection/File/8372/OFS_17-01_e.pdf
[3] While the effect of international oil prices on corporate financial performances is well documented in the literature (e.g. Henriques and Sadorsky, 2008, 2011; Dayanandan and Donker, 2011), the impact on the bank financial stability is poorly investigated.

Two different channels explain how oil prices affect the bank stability. The first one is the macroeconomic channel, which describes the link between oil prices and the macroeconomy, and how a degradation of macroeconomic conditions pass-through the banking sector performances. However, existing studies have examined this transmission channel without considering the different impact that oil prices may have on oil-exporting economies, as compared to oil-importing ones. For example, Kilian (2008) notes that oil price shocks negatively affect the consumption, and therefore the bank performances, through the uncertainty effect, precautionary savings effect, and the operating cost effect that lead to an increase of NPL. Nevertheless, in the case of an oil-exporting country, if oil prices increase at international level without recording a similar dynamic at national level, companies acting in the oil and gas industry, and the state, enregister higher revenues, with a positive effect on the banking sector (see for example Al-Khazali and Mirzaei, 2017). Therefore, the increase in international oil prices for oil-exporting countries does not necessary lead to higher production costs, reduction of purchasing power and economic growth contraction. On contrary, for oil-exporting countries a positive dynamic of oil prices might be associated with an increase of economic outcomes. In this case, banking performances improve (Demirgüç-Kunt and Huizinga, 2000; Athanasoglou et al., 2008).

Another element that should be considered inside the macroeconomic channel is the role of exchange rate. The link between oil prices and exchange rates is theoretically explained by Krugman (1983) and Golub (1983) who discuss the 'wealth effect channel', and by Amano and Van Norden (1998a, b), who describe the 'terms of trade channel'. It is generally accepted that a decrease of international oil prices leads to a depreciation of oil-exporting countries' currencies and *vice-versa* (Beckmann and Czudaj, 2013). In this line, a recent paper by Fedoseeva (2018) shows that the pass-through between oil prices and the Rubble exchange rate to US dollar substantially increased during the oil price collapse in 2014. Rubble's depreciation generated a sharp increase in import prices with a positive impact on inflation, threatening thus the banking sector stability. Moreover, the Bank of Russia implemented an inflation targeting (IT) strategy (Korhonen and Nuutilainen, 2017) and the oil pass-through inflation is considered to be higher for the IT countries (López-Villavicencio and Pourroy, 2019).

Finally, oil and gas revenues are very important for the fiscal stance of an oil-exporting country. As Malova and Van der Ploeg (2017) point in the case of Russia, if the chunk of oil and gas must be kept in the soil given the international agreements regarding global warming, or the lower level of international oil prices, the Russian fiscal stance needs to be tightened. In this case, the state may look for alternative financing sources and make appeal to public bank loans. Given the deterioration of the macroeconomic aggregates, a part of these loans may become non-performing, affecting thus the bank stability.

A second, unexploited channel, throughout international oil prices affect the bank performance and stability is the financial market channel. An overwhelming number of studies address the nexus between oil and stock markets, most of them documenting a positive correlation. The recent paper by Huang et al. (2017) investigates the nonlinearities in this relationship and shows that Russian stock market positively responds to the oil prices across all time scales. In this context, it is very likely that a decrease in oil prices will be correlated with a decrease in the share value of listed companies, including in the share value of listed public banks. Therefore, on the one hand the price to book value decreases, and hence the capacity of banks to generate sustainable earnings (Yildirim and Efthyvoulou, 2018). On the other hand, bank expend their lending and generate more income when the stock prices increase and during economic boom periods (Hesse and Poghosyan, 2009). Conversely, when stock prices decrease, the credit activity shrinks and the profitability decreases.

Against this background, we contribute to the existing literature in the following ways. First, we bring clarification to the oil price – bank stability relationship, making the distinction between the macroeconomic and financial market channel. At the same time, different from Lee and Lee (2018) who investigate the impact of oil prices on bank performances in China and document a negative relationship, we state that for an oil-exporting country as Russia, the macroeconomic impact of an increase in international oil prices is positive, and therefore, the impact on bank performance might also be positive. In fact, very few studies have assessed the effects of oil price changes on the bank financial stability in oil-dependent countries (e.g. Miyajima, 2016; Khandelwal et al., 2016). Our theoretical assumptions are similar with those advanced by IMF (2015), Husain et al. (2015) and Al-Khazali and Mirzaei (2017) who state that oil prices downturns adversely affect businesses in oil-rich economies, and therefore, the quality of bank loans. However, different from these studies we exploit both the macroeconomic and financial channels throughout the international oil prices affect the bank stability in Russia.

Second, in line with other studies assessing the determinants of bank stability (Laeven and Levine, 2009; Jeon and Lim, 2013; Lee and Hsieh, 2014; Kasman and Kasman, 2015) we use the Z-score as *proxy* for the bank stability.[4] However, different form previous works which mainly resort to the Boyd et al.'s (2006) approach to compute the Z-score, we also use an alternative measure, employing the Yeyati and Micco's (2007) methodology, for robustness purpose. The Z-score is a risk measure commonly used in the empirical banking literature to reflect banks' probability of insolvency and therefore, the level of bank stability. In addition, compared with the NPL ratio, data for Z-score computation is more accessible. Nevertheless, as Lepetit and Strobel (2013) show, there are several ways to compute the banking Z-score. Although there is a high correlation between these different metrics, the way the Z-score is computed may influence the empirical findings.

Third, different from previous works assessing the bank stability determinants and usually resort to General Method of Moments (GMM) models, we employ a Pool Mean Group (PMG) estimator. On the one hand, most of our series prove to be I(1) which makes the result of GMM estimators inconsistent. On the other hand, PMG exploits the cross-sectional dimension to gain more precise estimates of average long-run parameters and deals with the omitted variable bias. Both the long- and the short-run relationship between bank stability, international oil prices and the price to book value are estimated. In the short-run relationship, we include a series of control variables that may affect the bank stability. We consider the bank performance dimension and we include the net interest margins and the bank size (following Fungáčová and Poghosyan, 2011; Fina Katmani, 2018), the net operating cash flow (Beaver, 1968; Clark and Weinstein, 1983; Lanine and Vennet, 2006) and the liquidity rate (Fungáčová and Poghosyan, 2011; Lee and Lee, 2018). We address the role of bank competition and regulation quality as in Beck et al. (2013), Jeon and Lim (2013) and Kasman and Kasman (2015). We also consider the macroeconomic context including the GDP growth rate (Lee and Lee, 2018) and the institutional context, considering the perception of corruption (like Weill, 2011).

Forth, as far as we know this is the first study that investigates the role of shocks in international oil prices on the bank stability. Inspired by the studies of Hamilton (1996, 2003), Cong et al. (2008) and Babatunde et al. (2013), we compute both the positive and negative oil-price shocks and we test their impact on the bank stability. However, different from this works we propose an alternative approach, which allows to accommodate more shocks, and therefore to account for the effect of oil price volatility over a longer period. As in the case of reference methods, we use a rolling window to compute the shocks. Nevertheless, given our reduce time span (T = 9 for annual data) we compare the oil price level for a specific period with the average oil prices for the past periods, and not with their maximum/minimum values to identify positive/negative shocks.[5] This way we are able to identify the dispersion of an oil price shock over multiple periods. Consequently, we do not consider an oil shock as a sudden increase/decrease in oil prices and associate it with only one period. Instead, depending on the shock intensity, we are able to see its effects in time, over multiple periods.

---

[4] Data on NPL for public banks in Russia are in most of the cases unavailable. Therefore, the use of Z-scores represents a solution and a *proxy* for bank financial stability.

Fifth, we focus on 17 listed public banks from Russia, using FactSet data from 2008 to 2016 (annual data). There are several reasons for investigating the case of public banks sector. On the one hand, for Russia a decrease in international oil prices will first lead to a reduction of public exports, with a negative impact on the activity of public companies. Therefore, it is more likely that Russian public banks experience a deterioration of their financial performances and stability. On the other hand, different from other emerging markets economies, in Russia the market share of public banks is above 60% (Mamonov and Vernikov, 2017), which makes the public banks representative for the entire banking system. Finally, we consider the listed banks to test both the macroeconomic and the financial market channels throughout the oil prices influence the bank stability. The macroeconomic channel is more important for the public banks compared with the private ones, given the interaction between the public administration and public banks.

The remaining of the paper is as follows. Section 2 presents a brief literature review on bank stability determinants. Section 3 describes the data and the methodology. Section 4 shows the main empirical results while in Section 5 we present several robustness check analyses. The last section concludes.

## 2. Bank stability determinants: brief literature review

The literature assessing the determinants of bank stability is rather restrained. Most of the exiting works address the linkages between bank competition and financial stability. Two opposite views explain this relationship. First, the 'competition-fragility' theory states that an increased bank competition leads to a deterioration of bank stability (Keeley, 1990). According to this view, in a highly competitive environment, banks are determined to take more risks. Second, the 'competition-stability' theory argues that an increased competition improves the bank stability as contributes to a reduction of the information asymmetry, and to a decrease of the loan default rates. The empirical studies found evidence in the favor of both theories, while Martinez-Miera and Repullo (2010) provided a reconciliation of these opposite views, documenting a U-shaped relationship between competition and bank stability.

---

[5] Using a rolling window approach, Hamilton (2003) compares the oil price in the moment *t* with its maximum value over *n* previous periods to identify positive oil price shocks. Cong et al. (2008) compute both positive and negative oil price shocks but different from Hamilton, they identify oil price shocks by comparing the oil price in the moment *t* with its maximum/minimum values enregistered in all previous periods. Babatunde et al. (2013) combine these approaches and compute both positive and negative shocks, using in the same time a rolling window framework.

Recent empirical works on the subject continued, however, to report mixed findings. With a focus on Korean commercial banks and mutual savings banks, Jeon and Lim (2013) show that competition has a positive effect on the stability of mutual savings banks, a result in line with that reported by Boyd and De Nicolo (2005). Comparable results are obtained by Kasman and Kasman (2015) for the Turkish banking industry over the period 2002–2012. They show that competition is negatively related to the NPL ratio, and positively related to the Z-score, supporting the view of the 'competition-stability' theory. Opposite findings are reported by Fungáčová and Weill (2013) for a large sample of Russian banks. For the period preceding the 2009 global crisis, the authors support the view that a tighter bank competition enhances the occurrence of bank failures. In the same line, Beck et al. (2013) posit that competition has a larger impact on banks' fragility in highly regulated countries with developed stock markets.

The structure of bank ownership and its impact on the bank stability represents another focus point for the empirical literature. In this case also, the mixed findings might be supported by two concurrent hypotheses, derived from the foreign direct investment literature. Berger et al. (2000) differentiate between the 'global advantage' and 'home field advantage' hypotheses. In the first case, the foreign banks that invest abroad might have competitive advantages relative to their domestically-owned peers. These advantages include risk management skills and access to funds, fostering therefore their stability. The 'home field advantage hypothesis' affirm the opposite, showing that domestic banks are familiarized with the economic and institutional context, and this way can improve their financial performance. The recent study by Lee and Hsieh (2014) conducted for 1,387 individual banks in 27 selected Asian countries, covering the period 1995–2009, empirically investigate the relation between bank-level foreign ownership and stability. Relying on a GMM framework, the authors show mixed evidence. An inverted U-shaped relationship between foreign ownership and stability is documented.

A new strand of literature emerged during the last years, investigating the role of oil prices in explaining the bank stability. Building upon Husain et al. (2015), Al-Khazali and Mirzaei (2017) verify whether oil prices shocks have any impact on NPL ratio for 2,310 commercial banks in 30 oil-exporting countries over the period 2000–2014. The authors resort to a dynamic GMM model and show that a rise (fall) in international oil prices is associated with a decrease (increase) in NPL ratio. At the same time, the unfavorable impact of adverse oil prices dynamics on the quality of bank loans is more pronounced in the case of large banks. For a set of banks from the Gulf Cooperation Council (GCC) countries, Ibrahim (2019) underlines the favorable effects of positive oil price changes on bank profitability and credit growth, while underling the negative impact on NPL. Opposite results are reported by Lee and Lee (2018) for the Chinese banking sector. The authors assess bank performance through a broad array of CAMEL (Capital adequacy, Asset quality, Management, Earnings, and Liquidity) indicators, and discuss whether the correlations between oil prices and banks stability change with different dimensions of country risk. Their results reveal that oil prices negatively affect the bank performance. Adetutu et al. (2019) for Kazakhstan report a similar result and show that oil price booms negatively influence the banks' total productivity.

Different from these works, we exploit both the macroeconomic and the financial market channels to see how oil prices and bank valuation affect the level of bank stability. In addition, we assess the bank stability using different metrics for the Z-score. Further, we focus on the Russian banking sector which was severely affected by the recent crisis and on the listed public banks, to see how the international oil price dynamics and the currency depreciation influence their performance. Finally, we test the effect of oil price returns, but also the impact of positive/negative shocks in international oil prices.

## 3. Data and methodology

### *3.1. Data*

Data comes from the FactSet database, which initially include 20 listed public banks and cover a period from 2005 to 2017 (all data are expressed in US dollars). Out of the 20 banks, for three banks we have a very small amount of information[6] and two banks were privatized in 2017.[7] Given that our T is relatively small and we need a rolling window to compute the Z-score and oil price shocks, we have decided to cover the period 2008 to 2016, for 17 banks.[8] Therefore, the banks that were privatized in 2017 are included in the analysis. A rolling window of four years is used for the computation of moving means (n=4) for Z-scores. A higher Z-score is associated with an increased bank stability (the probability for a bank to make default decreases).

---

[6] For AK Bars Bank data are available starting with 2011, for RBC OJSC there are no data available for a series of indicators as liquidity ratio or net interest margins, and severe losses were recorded for the entire period. In addition, for the Best Efforts Bank data are available starting with 2014 only. These banks are therefore excluded from the analysis.

[7] These two banks are Promsvyazbank and Tatfondbank.

The general formula for the Z-score computation is (Albulescu and Ionescu, 2018):

$$\text{z-score}_t = \frac{CAR_t + ROA_t}{\sigma_{ROA,t}}, \qquad (1)$$

where CAR: represents the capital-to-assets ratio, ROA is the return on assets and $\sigma$ the standard deviation.

To compute the Z-score, we first, use the Boyd et al.'s (2006) approach (z1), which relies on the moving means $\mu_{CAR,t}(n)$, $\mu_{ROA,t}(n)$ and the standard deviation $\sigma_{ROA,t}(n)$, calculated for each period $t \in \{1 \ldots T\}$. Therefore, the formula become:

$$z1_t = \frac{\mu_{CAR,t} + \mu_{ROA,t}}{\sigma_{ROA,t}}. \qquad (2)$$

Second, for robustness purpose, we use the approach of Yeyati and Micco (2007), where the moving mean $\mu_{ROA,t}(n)$ and the standard deviation $\sigma_{ROA,t}(n)$ are calculated for each period $t \in \{1 \ldots T\}$, and are afterwards combined with the current value of $CAR_t$. Therefore, the formula for the Z-score (z2) become:

$$z2_t = \frac{CAR_t + \mu_{ROA,t}}{\sigma_{ROA,t}}. \qquad (3)$$

ROA is computed as the ratio between the net income and total assets. While the Z-score is the dependent variable, the main explanatory variables are represented by the oil prices[9] – wti, by the oil price positive ($wti^+$) and negative shocks ($wti^-$), and by the price to book value ratio (pbvr).

The oil price shocks for annual data are computed as follows[10]:

$$wti^+ = IF(wti_t > AVERAGE(wti_{t-1}:wti_{t-3}); wti_t - AVERAGE(wti_{t-1}:wti_{t-3}); 0). \quad (4)$$

$$wti^- = IF(wti_t < AVERAGE(wti_{t-1}:wti_{t-3}); wti_t - AVERAGE(wti_{t-1}:wti_{t-3}); 0). \quad (5)$$

---

[8] The 17 public banks retained in our sample are: Avangard Joint Stock Bank, Bank Otkritie Financial Corporation, Bank St. Petersburg, Bank Zenit, Credit Bank of Moscow, Far East Bank, Gazprombank, Joint Stock Commercial Bank Rosbank, Moscovskiy Oblastnoi Bank, OTP Bank, Bank Uralsib, Promsvyazbank, Sberbank Russia, Tatfondbank, Vozrozhdenie Bank, VTB Bank and West Siberian Commercial Bank.

[9] As in Lee and Lee (2018) we use WTI crude oil prices from the Energy Information Administration, expressed in log-returns.

[10] Figure A1 – Appendix shows how shocks spread over time.

The control variables first include the bank performances' dimension and are represented by the net interest margins – nim (FactSet), by net operating cash flow – nocf (FactSet), by the liquidity ratio – lr (FactSet) computed as the ratio between net loans and total deposits showing the maturity match (as in Lee and Lee, 2018) and by the size – ta, calculated as the natural log of total assets (FactSet). For all these variables we expect a positive influence on bank stability in the short run. The macroeconomic context is represented by the GDP growth rate – gdp (World Bank statistics) whereas the banking sector competition is assessed through the bank concentration index – bc (World Bank statistics). While the economic growth should have a positive influence on the stability level, the effect of bank competition is not straightforward (see 'competition-fragility' vs. 'competition-stability' theories). Finally, the quality of institutions is represented by the political risk associated with the regulatory quality[11] – rq (World Bank statistics) and by the Corruption Perception Index – cpi (Transparency International). A better regulatory framework should have a positive impact on bank stability, while the opposite applies for a higher corruption level.[12]

*3.2. Methodology*

The PMG estimator proposed by Pesaran et al. (1999) supposes an Autoregressive Distributive Lag (ARDL) framework, designed for dynamic panel specifications (ARDL $(p,q_1,\ldots,q_k)$):

$$z_{i,t} = \sum_{j=1}^{p} \lambda_{i,j} z_{i,t-j} + \sum_{j=0}^{q} \delta'_{i,j} X_{i,t-j} + \mu_i + \varepsilon_{i,t}, \qquad (6)$$

where: z is the Z-score, i is the number of groups (banks) and t is the number of periods (years), $X_{i,t}$ is the k×1 vector of explanatory variables, $\delta'_{i,j}$ are coefficients, $\lambda_{i,j}$ are scalars, $\mu_i$ are group effects, $\varepsilon_{i,t}$ is the error term.

If the variables are I(1), Eq. (6) can be reparametrized into an error correction equation, where additional control variables might be introduced (Blackburne and Frank, 2007):

$$\Delta z_{i,t} = \rho_i \left( z_{i,t-j} - \theta'_i X_{i,t} \right) + \sum_{j=1}^{p-1} \lambda^*_{i,j} \Delta z_{i,t-j} + \sum_{j=0}^{q-1} \delta^*_{i,j} \Delta X_{i,t-j} + \sum_{j=0}^{q-1} \gamma^*_{i,j} \Delta Y_{i,t-j} + \mu_i + \varepsilon_{i,t}, \qquad (7)$$

---

[11] The Russian banking regulation framework recorded important changes after the banking crisis in 2014. This element may also affect the bank stability. We consider that the World Bank indicator assessing the regulatory quality capture the effect of the banking regulation reform.
[12] The methodology used by Transparency International to assess the perception on corruption, associate a high value of cpi with a small level of corruption. Therefore, a positive sign for cpi is expected in our regressions.

where: $\rho_i$ is the error-correction speed of the adjustment term (which should be negative and significantly different from zero to validate the existence of a long-run relationship), $\theta_i$ is the vector that explains the long-run relationships between variables, $Y_{i,t}$ is the k×1 vector of control variables, $\lambda_{i,j}^*$, $\delta_{i,j}^*$ and $\gamma_{i,j}^*$ are short-run coefficients.

### *3.3. General statistics*

The summary statistics of our sample are presented in Table 1, showing a very high variability for the net operating cash-flow, but also for the Z-score. Further, the negative oil price shocks are higher that the positive ones over the analyzed time span. In addition, the bank competition considerably fluctuates over the analyzed period, from a minimum level of 21.58 to a maximum of 47.45.[13]

Table 1. Summary statistics

|  | z1 | z2 | wti | wti$^+$ | wti$^-$ | pbvr | nim | nocf | lr | ta | gdp | bc | rq | cpi |
|---|---|---|---|---|---|---|---|---|---|---|---|---|---|---|
| Mean | 22.70 | 22.17 | -0.057 | 8.273 | -11.43 | 0.798 | 45.04 | 43.04 | 1.148 | 4.596 | 1.149 | 31.66 | 0.697 | 2.544 |
| SD | 21.56 | 20.97 | 0.323 | 11.74 | 18.12 | 0.420 | 16.90 | 141.8 | 0.309 | 2.895 | 4.098 | 7.988 | 0.079 | 0.324 |
| Min | -0.631 | -2.182 | 0.320 | 34.66 | 0.000 | 1.642 | -9.549 | -36.74 | 0.452 | -0.982 | -7.820 | 47.56 | 0.773 | 2.100 |
| Max | 156.5 | 156.5 | -0.650 | 0.000 | -46.40 | 0.167 | 82.96 | 121.5 | 2.681 | 10.21 | 5.284 | 21.58 | 0.591 | 2.900 |

*Note: z1 – Z-score 1; z2 – Z-score 2; wti – WTI crude oil prices return; wti$^+$ – positive shocks in crude oil prices; wti$^-$ – negative shocks in crude oil prices; pbvr – price to book value ratio; nim – net interest margins; nocf – net operating cash flow; lr – liquidity ratio; ta – natural log of total assets; gdp – economic growth rate; bc – bank concentration; rq - regulatory quality; cpi – corruption perception index.*

We start our empirical analysis with a series of panel unit root tests. To see what generation of tests should be applied, we first test the existence of the cross-sectional independence hypothesis (Table 2). *A priori*, given the existence of the interbank market, it is hard to accept the independence hypothesis. However, all three tests (Frees, 1995; Friedman, 1937; Pesaran, 2004) do not reject the null hypothesis, showing that the first generation of panel unit root tests, which are more powerful in the presence of cross-sectional independence, should be applied.

Table 2. Cross-sectional dependence tests

---

[13] The correlation matrix (Table A1 – Appendix) shows a high correlation between the two metrics of the Z-score, namely z1 and z2. A positive correlation appears between bank stability and our interest variables, as expected. In addition, bank performances are positively correlated with the Z-score (nocf represents an exception and shows a negative correlation with the Z-score). At the same time, it seems that the size is positively correlated with Z-score, indicating that larger banks are more stable. Further, bank stability is positively correlated with the economic growth, as expected, but also with cpi (a higher cpi is equivalent with a lower perception of corruption). The level of correlation of our variables seems, however, reduced (except for the two metrics of the Z-score).

| Cross-sectional dependence tests | | | |
|---|---|---|---|
| *dependent variable* | Frees (1995) | Friedman (1937) | Pesaran (2004) |
| z1 | 0.795 | 13.08 | 0.417 |
| z2 | 0.688 | 12.73 | 0.706 |

*Note: \*\*\*,\*\*,\* means rejection of the null hypothesis of cross-sectional independence at 99%, 95% and 90% confidence level (equivalent with the existence of cross-sectional dependence).*

The panel unit root tests show mixed evidence (Table 3). While for z1, z2 and liquidity ratio, two out of for tests reject the null hypothesis of the presence of unit roots, only one test rejects the null hypothesis for the other variables. In the case of the regulation quality, bank competition and negative oil price shocks, all tests show the presence of unit roots. An exception is the gdp, where all tests indicates the presence of a mean reverting process. We may conclude that most of our variables are I(1), which recommend the PMG approach for estimation.

Table 3. Panel unit root tests

| | Levin, Lin & Chu t* | Im, Pesaran and Shin W-stat | ADF - Fisher Chi-square | PP - Fisher Chi-square |
|---|---|---|---|---|
| z1 | -3.459*** | -0.895 | 41.21 | 66.75*** |
| z2 | -4.292*** | -1.179 | 42.68 | 62.71*** |
| wti | -0.865 | -0.699 | 34.67 | 154.2*** |
| $wti^+$ | -0.771 | -0.262 | 29.26 | 147.6*** |
| $wti^-$ | 10.18 | 2.621 | 7.330 | 16.25 |
| pbvr | -3.658 | -0.829 | 45.43* | 37.46 |
| nim | 0.230 | 1.563 | 22.55 | 46.74* |
| nocf | 1.503 | 0.423 | 39.46 | 70.84*** |
| lr | -0.642*** | -0.140 | 38.77 | 103.1*** |
| ta | -5.170*** | -0.719 | 42.40 | 17.68 |
| gdp | -7.728*** | -2.196** | 55.63** | 97.72*** |
| bc | 5.433 | 4.571 | 2.316 | 14.83 |
| rq | -0.424 | 2.240 | 8.907 | 10.52 |
| cpi | -3.602*** | 1.794 | 11.07 | 9.877 |

*Notes: (i) the null hypothesis for all the tests is the presence of unit roots (the t\* test assumes common unit root process while the other tests assume individual unit root process);*
  *(ii) \*, \*\*, \*\*\*, mean stationarity (in level) significant at 10 %, 5 % and 1 %.*

## 4. Results

Applying the PMG estimator, we test nine different models. Given the characteristics of our sample with a small T and a reduced number of observations, we introduce in the short-run specification each control variable separately. Consequently, the long-run relationship between z, wti and pbvr remains the same for all models. The first model is run without control variables (Model 1). Afterwards, in the short-run equation we introduce separately the control variables, namely nim (Model 2), nocf (Model 3), lr (Model 4), ta (Model 5), gdp (Model 6), bc (Model 7), rq (model 8) and cpi (Model 6). We estimate these models for the oil prices returns – wti (Table 4), for the oil price positive shocks – wti$^+$ (Table 5) and for the oil price negative shocks – wti$^-$ (Table 6).

The first set of results is presented in Table 4. We may notice that for all models, except for Model 4, the long-run relationship between bank stability, oil prices and bank valuation is significant. Both an increase in international oil prices and in the value of banks perceived by the investors (i.e. share prices) compared to the book value (pbvr), positively influence the bank stability. Nevertheless, it seems that the direct, macroeconomic channel that explains the oil-bank stability pass-through is more important compared with the financial channel in the case of Russian public banks.

These results are in agreement with those reported by Khandelwal et al. (2016), showing that a decline in oil prices leads to an increase in the NPL ratio (that is, the bank instability). The adjustment coefficient from the short-run equation ($\rho_i$) is negative and significant in all cases, providing evidence in the favor of a significant long-run relationship. However, when we look to the short-run coefficients, we observe that in almost all the cases these coefficients are not significant, except for the Model 4, where a positive relationship appears in the short run between the liquidity ratio and bank stability. In fact, as Khandelwal et al. (2016) emphasize, in the short run the effect of oil prices might be captured by the macroeconomic variables (e.g. growth rate), which explain the loss of the significance of oil prices coefficients.

Table 4. Results for the oil price returns – main (z1)

|  | z1 | Model 1 | Model 2 | Model 3 | Model 4 | Model 5 | Model 6 | Model 7 | Model 8 | Model 9 |
|---|---|---|---|---|---|---|---|---|---|---|
| Long-run | wti | 10.50*** | 11.48*** | 15.14*** | 4.077 | 17.47*** | 10.22*** | 9.832*** | 13.74*** | 14.62*** |
|  | pbvr | 0.191*** | 0.195*** | 0.179*** | 0.186*** | 0.194*** | 0.183*** | 0.192*** | 0.196*** | 0.166*** |
| Short-run | $\rho_i$ | -0.797*** | -0.911*** | -0.819*** | -0.947*** | -0.720*** | -0.729*** | -0.782*** | -0.808*** | -0.751*** |
|  | wti | -2.245 | 11.58 | -1.363 | -7.405* | -2.084 | 4.016 | -3.296 | -5.589 | -4.431 |
|  | pbvr | 13.39 | -88.69 | 79.84 | -111.7 | 19.68 | 158.2 | 1.367 | 15.27 | 217.8 |
|  | nim |  | -0.359 |  |  |  |  |  |  |  |
|  | nocf |  |  | -6.548 |  |  |  |  |  |  |
|  | lr |  |  |  | 23.05** |  |  |  |  |  |
|  | ta |  |  |  |  | 6.377 |  |  |  |  |
|  | gdp |  |  |  |  |  | 0.230 |  |  |  |
|  | bc |  |  |  |  |  |  | -0.116 |  |  |
|  | rq |  |  |  |  |  |  |  | 4.321 |  |
|  | cpi |  |  |  |  |  |  |  |  | 9.713 |
|  | c | 7.713 | 5.041 | 1.140 | 13.16* | 9.202 | 12.97* | 6.972 | 6.749 | 10.87 |
| Log Likelihood |  | -453.3 | -430.3 | -421.1 | -432.6 | -418.3 | -429.1 | -440.3 | -425.6 | -423.2 |

Note: ***, **, * indicates significance at 1%, 5% and 10%.

In what follows we compare the effect of positive and negative oil price shocks on the bank stability. Table 5 shows that $wti^+$ have a positive long-run impact on the bank stability, in all the cases (again, Model 4 represent an exception).[14] However, the effect of oil price shocks on bank stability is much more reduced compared to the effect of price returns. Notice that the approach used for the shocks' computation allows the propagation of shocks in time. Therefore, the effect of a price shock can be recorder over one, two, or even three consecutive periods. These findings show that not the shock itself is important for the bank stability, but the increase in the oil price associated with a positive shock. It is also interesting to notice the sign of $wti^+$ in the short-run equation, which is negative. Although these coefficients are not significant (Models 4 and 9 represents an exception), they suggest that the short-run impact of the shock is negative, result that explains the findings of Adetutu et al. (2019), stating that an oil price boom negatively affects the bank performances in Kazakhstan, which similar to Russia, represents an oil-exporting country. Indeed, in the short-run, the domestic consumption slows down following a sudden increase in oil prices, which in turn negatively affects companies' performances and their capacity to fulfil their financial obligations. However, in the long run, the shock is absorbed and, bringing benefits for the Russian economy, it contributes to the bank stability by raising the revenues of the state companies and ameliorating the fiscal stance.

Table 5. Results for the oil price positive shocks – main (z1)

|  | z1 | Model 1 | Model 2 | Model 3 | Model 4 | Model 5 | Model 6 | Model 7 | Model 8 | Model 9 |
|---|---|---|---|---|---|---|---|---|---|---|
| Long-run | $wti^+$ | 0.114*** | 0.112*** | 0.749*** | -0.045 | 0.187*** | 0.153*** | 0.093*** | 0.142*** | 2.008*** |
|  | pbvr | 0.212*** | 0.217*** | 0.109*** | 0.177*** | 0.226*** | 0.218*** | 0.587*** | 0.076*** | 0.099*** |
| Short-run | $\rho_i$ | -0.825*** | -0.784*** | -0.646*** | -1.033*** | -0.799*** | -0.772*** | -0.554*** | -0.892*** | -0.628*** |
|  | $wti^+$ | -0.092 | -0.338 | -0.219 | -0.461** | -0.084 | -0.226 | 0.118 | -0.052 | -0.299** |
|  | pbvr | -13.58 | -30.83 | 0.838 | -28.42 | 79.52 | 13.90 | -15.83 | 22.78 | 97.79 |
|  | nim |  | 0.584 |  |  |  |  |  |  |  |
|  | nocf |  |  | -5.565 |  |  |  |  |  |  |
|  | lr |  |  |  | 33.97** |  |  |  |  |  |
|  | ta |  |  |  |  | 10.74 |  |  |  |  |
|  | gdp |  |  |  |  |  | 0.242 |  |  |  |
|  | bc |  |  |  |  |  |  | 0.111 |  |  |
|  | rq |  |  |  |  |  |  |  | 7.774** |  |
|  | cpi |  |  |  |  |  |  |  |  | 3.528* |
|  | c | 6.969 | -6.568 | -4.671 | 9.447 | 7.496 | 4.917 | 13.47 | 10.56* | 1.815 |
| Log Likelihood |  | -457.1 | -424.7 | -449.0 | -426.8 | -429.1 | -427.0 | -430.6 | -416.9 | -427.1 |

Note: ***, **, * indicates significance at 1%, 5% and 10%.

As in the previous case (Table 4), a positive bank valuation (higher price to book ratio) contributes to the bank stability. This result shows that positive oil price shocks induce an indirect positive effect on the stock market, which contribute to a better bank valuation. Different from the previous results, for Models 8 and 9 the coefficient of the control variables is significant and have the expected sign, showing that both regulatory quality and a reduced perception of corruption favorize the stability of the public banks.

---
[14] The impact of the liquidity ratio might capture the effect of international oil prices.

Table 6 presents the results for negative shocks in oil prices. As expected, the long-run impact is negative and significant (except for the Models 4 and 8). The pbvr has a long-run positive influence on bank stability, for all tested models. In this case, the economic growth rate has a slight influence on the short-run bank stability (Model 6). It appears that in the short run, the initially impact of a negative shock is positive (Models 5 and 6). However, in the long run, the results clearly show the opposite.

Table 6. Results for the oil price negative shocks – main (z1)

|  | z1 | Model 1 | Model 2 | Model 3 | Model 4 | Model 5 | Model 6 | Model 7 | Model 8 | Model 9 |
|---|---|---|---|---|---|---|---|---|---|---|
| Long-run | $wti^-$ | -0.054*** | -0.131*** | -0.069*** | -0.025 | -0.227*** | -0.147*** | -0.059*** | -0.040 | -0.028*** |
|  | pbvr | 0.194*** | 0.015*** | 0.188*** | 0.176*** | 0.009* | 0.035*** | 0.187*** | 0.063*** | 0.068*** |
| Short-run | $\rho_i$ | -0.755*** | -0.584*** | -0.726*** | -0.981*** | -0.452*** | -0.612*** | -0.743*** | -0.755*** | -0.816*** |
|  | $wti^-$ | 0.098 | -0.113 | 0.101 | 0.017 | 0.224** | 0.573* | 0.255 | 0.125 | 0.123 |
|  | pbvr | 5.397 | -56.20 | 126.1 | -22.45 | -168.8 | -74.32 | 57.18 | 6.600 | 144.4 |
|  | nim |  | -0.374 |  |  |  |  |  |  |  |
|  | nocf |  |  | -3.795 |  |  |  |  |  |  |
|  | lr |  |  |  | 21.52* |  |  |  |  |  |
|  | ta |  |  |  |  | 1.523 |  |  |  |  |
|  | gdp |  |  |  |  |  | 1.053* |  |  |  |
|  | bc |  |  |  |  |  |  | 0.504 |  |  |
|  | rq |  |  |  |  |  |  |  | 6.859* |  |
|  | cpi |  |  |  |  |  |  |  |  | 7.120 |
|  | c | 6.561 | 4.350 | 1.225 | 12.03 | 7.326* | 3.042 | 6.476 | 8.040 | 9.048 |
| Log Likelihood |  | -460.6 | -433.6 | -435.1 | -424.1 | -441.3 | -439.6 | -443.0 | -439.5 | -420.1 |

Note: ***, **, * indicates significance at 1%, 5% and 10%.

These findings confirms the previous results reported in the literature (e.g. Al-Khazali and Mirzaei, 2017; Ibrahim, 2019), providing support for asymmetric effects of oil price shocks. However, different from Ibrahim (2019), we show that both positive and negative oil price changes influence bank performance. In addition, different from Al-Khazali and Mirzaei (2017) we posit that oil price positive shocks have a greater impact on the bank stability compared with negative shocks. Long-run expectations related to oil price positive jumps might determine banks to increase the loans' volume in a favorable economic context, which allows them to make more profit and to strengthen their financial stability.

Nevertheless, our estimations might be subject to some caveats given the lack of significance for the control variables' coefficients in the short-run equation for most of the tested models, situation that requires additional investigations. Therefore, in the next section we perform two different robustness check analyses. First, we use an alternative measure for the Z-score, relying on the Yeyati and Micco's (2007) approach. Second, we drop from our data sample the Sberbank Russia statistics. Sberbank might me considered as an outlier in our sample, being the largest bank in the Russian banking sector. Its assets in 2016 represents about 50% from the total public bank system.

## 5. Robustness analysis

### 5.1. Alternative measure for the Z-score

The robustness results using z2 are presented in Table 7. Similar to the main analysis, we document the existence of a long-run relationship between bank stability on the one hand, and dynamics of oil prices and bank valuation on the other hand. Both explanatory variables positively influence the stability of Russian public banks in the long run. A slight difference appears in the case of Models 6 and 8, where the sign of the oil price coefficient is either insignificant, or negative. Similar to the main results, in the short run there is no significant influence of oil prices on the bank stability, while the coefficients of control variables are not significant (except for Model 4).

Table 7. Results for the oil price returns – robustness (z2)

|  | z2 | Model 1 | Model 2 | Model 3 | Model 4 | Model 5 | Model 6 | Model 7 | Model 8 | Model 9 |
|---|---|---|---|---|---|---|---|---|---|---|
| Long-run | wti | 5.888* | 8.153*** | 18.40*** | 1.829 | 10.93*** | 3.107 | 2.653*** | -1.791 | 14.19*** |
|  | pbvr | 0.147*** | 0.159*** | 0.114*** | 0.130*** | 0.147*** | 0.105*** | 0.152*** | 0.107*** | 0.102*** |
| Short-run | $\rho_i$ | -0.773*** | -0.897*** | -0.793*** | -0.893*** | -0.684*** | -0.660*** | -0.741*** | -0.716*** | -0.937*** |
|  | wti | -1.748 | 9.130 | -3.812 | -1.489 | 0.216 | 4.357 | -3.478 | -1.296 | -13.23 |
|  | pbvr | 81.76 | 19.61 | 183.6 | -72.91 | 45.89 | 22.58 | 75.07 | 81.48 | 33.05 |
|  | nim |  | 0.303 |  |  |  |  |  |  |  |
|  | nocf |  |  | -6.091 |  |  |  |  |  |  |
|  | lr |  |  |  | 31.23*** |  |  |  |  |  |
|  | ta |  |  |  |  | -9.482 |  |  |  |  |
|  | gdp |  |  |  |  |  | -0.212 |  |  |  |
|  | bc |  |  |  |  |  |  | -0.236 |  |  |
|  | rq |  |  |  |  |  |  |  | -1.927 |  |
|  | cpi |  |  |  |  |  |  |  |  | 2.107 |
|  | c | 7.726 | 5.558 | 3.042 | 14.30* | 9.049 | 12.45** | 6.334 | 6.808 | 10.88* |
| Log Likelihood |  | -463.5 | -439.9 | -429.7 | -432.0 | -440.7 | -433.0 | -438.1 | -431.2 | -427.7 |

Note: ***, **, * indicates significance at 1%, 5% and 10%.

The robustness results related to the impact of positive and negative oil price shocks on bank stability are presented in Appendix. Table A2 shows that positive shocks have a long run and benefic effect on the bank stability, and confirm thus the main findings. The same applies for the effects of negative shocks (Table A3), which contribute to a reduction of the stability level in the long run (except for Model 5). The effect of the control variables is practically insignificant in the short-run equation. This robustness analysis confirms the previous findings, showing the absence of a significant effect of oil price shocks on the bank stability in the short run.

*5.2. Re-sampling results (16 cross-sections)*

The second robustness check implies the construction of a new dataset, including 16 banks, for the same period 2008 to 2016 (these results are presented in Table 8). Although the long-run influence of international oil prices seems to be less important if we compare the level of coefficients, it remains positive and significant (the significance vanishes in this case for Model 8 only). At the same time, the bank positive valuation by investors contribute to increasing bank stability in the long run. Nevertheless, in the short run, no significant influence is recorded. This evidence confirms the main findings and state that the influence of oil prices on bank stability can be documented only in the long run.

If we now refer to the impact of oil price shocks, Tables A4 and A5 (Appendix) show that positive oil price shocks enhance bank stability in the long run (except for Model 8), while negative shocks have an opposite effect, as expected. Like in the main analysis, we notice that for Models 5 and 6, the short-run impact of a negative shock is positive, being associated with a cost reduction for households and companies, which favor the consumption and investment. However, in the long run, the results clearly indicate the negative impact in international oil price drops, on the Russian public banks stability.

Table 8. Results for the oil price returns – robustness (re-sampling)

|  | z1 | Model 1 | Model 2 | Model 3 | Model 4 | Model 5 | Model 6 | Model 7 | Model 8 | Model 9 |
|---|---|---|---|---|---|---|---|---|---|---|
| Long-run | wti | 5.240* | 12.53*** | 17.57*** | -40.46 | 9.890** | 14.30** | 4.343* | -3.711 | -9.239* |
|  | pbvr | 0.514*** | 0.098*** | 0.114*** | 0.750*** | 0.423*** | 0.088*** | 0.094*** | 0.076*** | 0.110*** |
|  | $\rho_i$ | -0.573*** | -0.861*** | -0.712*** | -0.660*** | -0.564*** | -0.632*** | -0.691*** | -0.759*** | -0.692*** |
|  | wti | -0.017 | 13.44 | -1.854 | 4.613 | 1.425 | 3.629 | -3.701 | 1.496 | 3.874 |
|  | pbvr | 15.99 | -91.04 | 88.47 | -15.05 | -10.90 | 16.55 | 2.204 | 23.36 | 16.81 |
|  | nim |  | 0.401 |  |  |  |  |  |  |  |
|  | nocf |  |  | -6.689 |  |  |  |  |  |  |
| Short-run | lr |  |  |  | 34.52** |  |  |  |  |  |
|  | ta |  |  |  |  | 8.053 |  |  |  |  |
|  | gdp |  |  |  |  |  | -0.236 |  |  |  |
|  | bc |  |  |  |  |  |  | -0.265 |  |  |

| | | | | | | | | | |
|---|---|---|---|---|---|---|---|---|---|
| rq | | | | | | | | -6.310 | |
| cpi | | | | | | | | | 11.23 |
| c | 7.975 | 5.740 | -0.126 | 15.68* | 8.672 | 12.64** | 5.203 | 7.441 | 8.562 |
| Log Likelihood | -436.1 | -424.1 | -414.0 | -397.2 | -417.4 | -417.8 | -432.2 | -415.6 | -409.1 |

Note: ***, **, * indicates significance at 1%, 5% and 10%.

In a nutshell, we reinforce the findings by Khandelwal et al. (2016), Al-Khazali and Mirzaie (2017) and Ibrahim (2019), reporting a significant and positive effect of oil prices on bank stability in Russia. However, different from these previous findings, we show that the oil price-bank stability pass-through is only significant in the long run, whereas the positive shocks in oil prices have a larger influence on bank stability compared with negative shocks.

# 6. Conclusions

This paper adds to the literature investigating the determinants of bank stability, with a focus on the role of international oil prices. For this purpose, complementary to previous works we exploit two channels throughout the oil prices dynamics may influence the bank stability, namely the macroeconomic and the financial market channel. Different from previous studies on the subject, we test not only the influence of international oil price returns, but also the effect of positive and negative shocks in oil prices, on the bank stability.

To this end, we use a PMG estimator for a sample of 17 listed public banks from Russia, for the period 2008 to 2016. We show that both international oil prices and price to book value ratio have a positive impact on bank stability. Our results are in line with those reported by Husain et al. (2015) and Al-Khazali and Mirzaei (2017) which use a different proxy for the bank stability in oil-exporting countries. Nevertheless, our findings bring additional insights to the literature, showing that oil price shocks have a different impact on the stability in the short run, compared to the long run. Further, our results state that not only the macroeconomic, direct channel is important for oil price pass-through bank stability, but also the indirect, financial channel. These findings are supported by the robustness analyses we perform and show, once again, the importance of the oil prices volatility for the Russian economy. The findings clearly underline the positive long run effect of an increase in international oil price on bank stability in an oil-exporting country.

Two policy implications result from our investigation. First, it is important to disentangle between the short-run and long-run effects of international oil prices on bank stability. Positive oil price shocks may contribute to better bank performances in the long run, while having an opposite effect in the short run. The same applies for the negative shocks, although the short run influence is rather insignificant. Second, the authorities from oil-exporting countries should be aware by the fact that a positive oil production shock generate a negative oil price shock on the international market, with a negative influence on the domestic banking sector stability. Therefore, it is recommended to control the production, to obtain a smooth increase of international oil prices. This strategy helps the authorities to safeguard the public bank stability and, given the importance of public banks for the Russian financial sector, the stability of the entire financial system. The control of oil production by the Russian authorities is possible given the higher concentration of this sector, where the state-own company Rosneft accounts for nearly half of Russia's oil production (Simola and Solnnko, 2017). However, the Russian authorities did not tried to influence the international oil prices after 2014 by reducing the oil production.

Our results should be, however, interpreted with caution, and require additional investigations. On the one hand, in the short run, the control variables we use do not explain the level bank stability and are rather insignificant. On the other hand, our sample is relatively small, and do not allow the comparison between oil price effect on public and private banks. Finally, constrained by the length of our data sample, we use a linear framework, for a period characterized by important economic events for the economy of Russia.

## References

Adetutu, M.O., Odusanya, K.A., Ebireri, J.E. and Murinde, V. 2019. Oil booms, bank productivity and natural resource curse in finance. Economics Letters. https://doi.org/10.1016/j.econlet.2019.07.002.

Ahamed, M.M. and Mallick, S. 2017. Does regulatory forbearance matter for bank stability? Evidence from creditors' perspective. Journal of Financial Stability, 28, 163–180.

Albulescu, C.T. and Ionescu, A.M. 2018. The long-run impact of monetary policy uncertainty and banking stability on inward FDI in EU countries. Research in International Business and Finance, 45, 72–81.

Al-Khazali, O.M. and Mirzaei, A. 2017. The impact of oil price movements on bank non-performing loans: Global evidence from oil-exporting countries. Emerging Markets Review, 31, 193–208.

Amano, R. and Van Norden, S. 1998a. Exchange rates and oil prices. Review of International Economics, 6, 683–694.

Amano, R. and Van Norden, S. 1998b. Oil prices and the rise and fall of the US real exchange rate. Journal of International Money and Finance, 17, 299–316.

Athanasoglou, P.P., Brissimis, S.N. and Delis, M.D. 2008. Bank-specific, industry-specific and macroeconomic determinants of bank profitability. Journal of International Financial Markets, Institutions and Money, 18, 121–136.

Babatunde, M.A., Adenikinju, O., Adenikinju, A.F., 2013. Oil price shocks and stock market behaviour in Nigeria. Journal of Economic Studies, 40, 180–202.

Beaver, W. 1968. Market prices, financial ratios and the prediction of failure. Journal of Accounting Research, 6, 170–192.

Beck, T., De Jonghe, O. and Schepens, G. 2013. Bank competition and stability: Cross-country heterogeneity. Journal of Financial Intermediation, 22, 218–244.

Beckmann, J. and Czudaj, R., 2013. Is there a homogeneous causality pattern between oil prices and currencies of oil importers and exporters? Energy Economics, 40, 665–678.

Berger, A.N., DeYoung, R., Genay, H. and Udell, G. 2000. Globalisation of Financial Institutions: Evidence from Cross-border Banking Performance. Brookings–Wharton Papers on Financial Services, 3, 23–158.

Blackburne III E.F. and Frank, M.W. 2007. Estimation of nonstationary heterogeneous panels. Stata Journal, 7, 197–208.

BP. 2016. Statistical indicators. BP, London.

Boyd, J. and De Nicolo, G. 2005. The theory of bank risk—taking and competition revisited. Journal of Finance, 3, 1329–1343.

Boyd, J., De Nicoló, G. and Jalal, A. 2006. Bank risk-taking and competition revisited: new theory and new evidence. IMF Working Paper 06/297.

Clark, T. and Weinstein, M. 1983. The behavior of common stocks of bankrupt firms. Journal of Finance, 38, 489–504.

Cong, R.-G., Wei, Y.-M., Jiao, J.-L., Fan, Y., 2008. Relationships between oil price shocks and stock market: an empirical analysis from China. Energy Policy, 36, 3544–3553.

Dayanandan, A., and Donker, H. 2011. Oil prices and accounting profits of oil and gas companies. International Review of Financial Analysis, 20, 252–257.

De Jonghe, O. 2010. Back to the basics in banking? a micro-analysis of banking system stability. Journal of Financial Intermediation, 19, 387–417.

Demirgüç-Kunt, A. and Detragiache, E. 2000. Monitoring banking sector fragility: a multivariate logit approach. World Bank Economic Review, 14, 287–307.

Dreger, C., Kholodilin, K.A., Ulbricht, D. and Fidrmuc, J. 2016. Between the hammer and the anvil: The impact of economic sanctions and oil prices on Russia's ruble. Journal of Comparative Economics, 44, 295–308.

Duport, N., Fina, É., Goyeau, D. 2018. Diversification des institutions financières et risque systémique : la prise en compte des risques extrêmes. Revue Economique 69, 477–504.

Eller, M., Fidrmuc, J. and Fungáčová, Z. 2016. Fiscal policy and regional output volatility: evidence from Russia. Regional Studies, 50, 1849–1862.

Fedoseeva, S. 2018. Under pressure: Dynamic pass-through of oil prices to the RUB/USD exchange rate. International Economics, 156, 117–126.

Fernández, A.I., González, F. and Suárez, N. 2016. Banking stability, competition, and economic volatility. Journal of Financial Stability, 22, 101–120.

Fina Katmani, E. 2018. The effect of non-traditional banking activities on systemic risk: Does bank size matter? Finance Research Letters. https://doi.org/10.1016/j.frl.2018.10.013.

Frees, E. W. 1995. Assessing cross-sectional correlations in panel data. Journal of Econometrics, 69, 393–414.

Friedman, M. 1937. The use of ranks to avoid the assumption of normality implicit in the analysis of variance. Journal of the American Statistical Association, 32, 675–701.

Fungáčová, Z. and Poghosyan, T. 2011. Determinants of bank interest margins in Russia: Does bank ownership matter? Economic Systems, 35, 481–495.

Fungáčová, Z. and Weill, L. 2013. Does competition influence bank failures? Evidence from Russia. Economics of Transition, 21, 301–322.

García-Kuhnert, Y., Marchica, M.-T. and Mura, R. 2015. Shareholder diversification and bank risk-taking. Journal of Financial Intermediation, 24, 602–635.

Golub, S. 1983. Oil prices and exchange rates. Economic Journal, 93, 576–593.

Hamilton, J.D. 1996. This is what happened to the oil price-macroeconomy relationship. Journal of Monetary Economics, 38, 215–220.

Hamilton, J.D. 2003. What is an oil shock? Journal of Economics, 113, 363–398.

Henriques, I. and Sadorsky, P. 2008. Oil prices and the stock prices of alternative energy companies. Energy Economics, 30, 998–1010.

Henriques, I. and Sadorsky, P. 2011. The effect of oil price volatility on strategic investment. Energy Economics, 33, 79–87.

Hesse, H. and Poghosyan, T. 2016. Oil prices and bank profitability: evidence from major oil exporting countries in the Middle East and North Africa. IMF Working Paper No. 09/220.

Huang, S., An, H., Gao, X., Wen, S. and Hao, X. 2017. The multiscale impact of exchange rates on the oil-stock nexus: Evidence from China and Russia. Applied Energy, 194, 667–678.

Husain, A.M., Arezki, R., Breuer, P., Haksar, H., Helbling, T., Medas, P.A. and Sommer, M. 2015. Global implications of lower oil prices. IMF Staff Discussion Note, SDN/15/15.

Ibrahim, M.H. 2019. Oil and macro-financial linkages: Evidence from the GCC countries. The Quarterly Review of Economics and Finance, 72, 1–13.

IMF. 2015. Oil prices, financial stability, and the use of countercyclical macroprudential policies in the GCC. Annual Meeting Doha Qatar.

Jeon, J.Q. and Lim, K.K. 2013. Bank competition and financial stability: A comparison of commercial banks and mutual savings banks in Korea. Pacific-Basin Finance Journal, 25, 253–272.

Kasman, S. and Kasman, A. 2015. Bank competition, concentration and financial stability in the Turkish banking industry. Economic Systems, 39, 502–517.

Khandelwal, P., Miyajima, K. and Santos, A.O. 2016. The impact of oil prices on the banking system in the GCC, IMF Working Paper, WP/16/161.

Keeley, M.C. 1990. Deposit insurance, risk, and market power in banking. American Economic Review, 80, 1183–1200.

Kilian, L. 2008. The economic effects of energy price shocks. Journal of Economic Literature, 46, 871–909.

Korhonen, I. and Nuutilainen, R. 2017. Breaking monetary policy rules in Russia. Russian Journal of Economics, 3, 366–378.

Krugman, P., 1983. Oil and the dollar. NBER Working Paper No. 554.

Laeven, L. and Levine, R. 2009. Bank governance, regulation and risk taking. Journal of Financial Economics, 93, 259–275.

Lanine, G. and Vennet, R.V. 2006. Failure prediction in the Russian bank sector with logit and trait recognition models. Expert Systems with Applications, 30, 463–478.

Lee, C.-C. and Hsieh, M.-F. 2014. Bank reforms, foreign ownership, and financial stability. Journal of International Money and Finance, 40, 204–224.

Lee, C.-C. and Lee, C.-C. 2018. Oil price shocks and Chinese banking performance: Do country risks matter? Energy Economics. https://doi.org/10.1016/j.eneco.2018.01.010

Lepetit, L. and Strobel, F. 2013. Bank insolvency risk and time-varying Z-score measures. Journal of International Financial Markets, Institutions and Money, 25, 73–87.

López-Villavicencio, A. and Pourroy, M. 2019. Inflation target and (a)symmetries in the oil price pass-through to inflation. Energy Economics, 80, 860–875.

Mamonov, M. and Vernikov, A. 2017. Bank ownership and cost efficiency: New empirical evidence from Russia. Economic Systems, 41, 305–319.

Martinez-Miera, D., Repullo, R. 2010. Does competition reduce the risk of bank failure? Review of Financial Studies, 23, 3638–3664.

Miyajima, K. 2016. An empirical investigation of oil-macro-financial linkages in Saudi Arabia. IMF Working Paper, WP/16/22.

Pesaran, M.H. 2004. General diagnostic tests for cross section dependence in panels. University of Cambridge, Faculty of Economics, Cambridge Working Papers in Economics No. 0435.

Pesaran, M.H., Shin, Y. and Smith, R.P. 1999. Pooled mean group estimation of dynamic heterogeneous panels. Journal of American Statistics Association, 94, 621–634.

Simola, H. and Solanko, L. 2017. Overview of Russia's oil and gas sector. BOFIT Policy Brief, No. 5.

Tuzova, Y. and Qayum, F. 2016. Global oil glut and sanctions: The impact on Putin's Russia. Energy Policy, 90,140–151.

Wagner, W. 2010. Diversification at financial institutions and systemic crises. Journal of Financial Intermediation, 19, 373–386.

Weill, L. 2011. How corruption affects bank lending in Russia. Economic Systems, 35, 230–243.

Yeyati, E.L. and Micco, A. 2007. Concentration and foreign penetration in Latin American banking sectors: impact on competition and risk. Journal of Banking & Finance, 31, 1633–1647.

Yildirim, C. and Efthyvoulou, G. 2018. Bank value and geographic diversification: regional vs global. Journal of Financial Stability, 36, 225–245.

https://www.cbr.ru/Collection/Collection/File/8376/fin-stab-2014-15_4-1_e.pdf.

https://www.cbr.ru/Collection/Collection/File/8372/OFS_17-01_e.pdf.

# Appendix

Figure A1. Oil price shocks

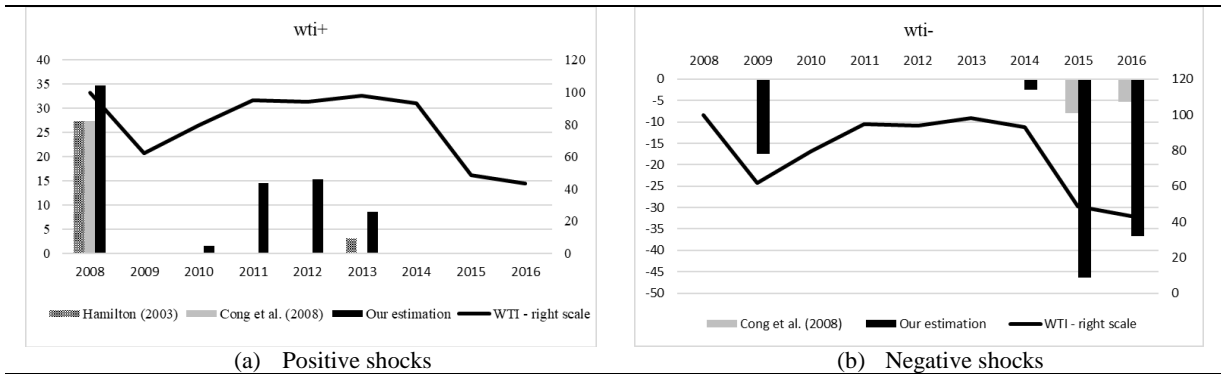

(a) Positive shocks  (b) Negative shocks

*Note: Hamilton's (2003) approach is designed to compute positive shocks only.*

Table A1. Correlation matrix

|       | z1     | z2     | wti    | wti$^+$ | wti$^-$ | pbvr   | nim    | nocf   | lr     | ta     | gdp    | bc     | rq     | cpi   |
|-------|--------|--------|--------|---------|---------|--------|--------|--------|--------|--------|--------|--------|--------|-------|
| z1    | 1.000  |        |        |         |         |        |        |        |        |        |        |        |        |       |
| z2    | 0.990  | 1.000  |        |         |         |        |        |        |        |        |        |        |        |       |
| wti   | 0.286  | 0.284  | 1.000  |         |         |        |        |        |        |        |        |        |        |       |
| wti$^+$ | 0.133 | 0.122 | 0.061 | 1.000   |         |        |        |        |        |        |        |        |        |       |
| wti$^-$ | -0.092 | -0.077 | -0.072 | -0.049 | 1.000  |        |        |        |        |        |        |        |        |       |
| pbvr  | 0.250  | 0.240  | 0.020  | 0.044   | -0.026  | 1.000  |        |        |        |        |        |        |        |       |
| nim   | 0.117  | 0.146  | 0.277  | 0.246   | 0.326   | -0.013 | 1.000  |        |        |        |        |        |        |       |
| nocf  | -0.018 | -0.015 | -0.051 | -0.088  | 0.096   | -0.058 | 0.151  | 1.000  |        |        |        |        |        |       |
| lr    | 0.200  | 0.275  | 0.207  | 0.349   | 0.192   | 0.130  | -0.121 | -0.017 | 1.000  |        |        |        |        |       |
| ta    | 0.101  | 0.089  | -0.034 | -0.062  | -0.052  | 0.237  | 0.149  | 0.490  | 0.131  | 1.000  |        |        |        |       |
| gdp   | 0.129  | 0.120  | 0.902  | 0.615   | -0.624  | 0.004  | 0.252  | -0.046 | 0.152  | 0.008  | 1.000  |        |        |       |
| bc    | -0.217 | -0.216 | -0.634 | -0.633  | -0.907  | 0.060  | -0.325 | 0.115  | -0.246 | 0.088  | -0.456 | 1.000  |        |       |
| rq    | 0.052  | 0.062  | 0330   | 0.470   | 0.533   | -0.078 | 0.229  | -0.122 | 0.128  | -0.106 | 0.256  | -0.759 | 1.000  |       |
| cpi   | 0.011  | 0.012  | -0.391 | -0.374  | 0.048   | 0.095  | -0.183 | 0.119  | -0.168 | 0.116  | -0.166 | 0.715  | -0.869 | 1.000 |

*Note: z1 – Z-score 1; z2 – Z-score 2; wti – WTI crude oil prices return; wti$^+$ – positive shocks in crude oil prices; wti$^-$ – negative shocks in crude oil prices; pbvr – price to book value ratio; nim – net interest margins; nocf – net operating cash flow; lr – liquidity ratio; ta – natural log of total assets; gdp – economic growth rate; bc – bank concentration; rq – regulatory quality; cpi – corruption perception index.*

Table A2. Results for the oil price positive shocks – robustness (z2)

|  | z2 | Model 1 | Model 2 | Model 3 | Model 4 | Model 5 | Model 6 | Model 7 | Model 8 | Model 9 |
|---|---|---------|---------|---------|---------|---------|---------|---------|---------|---------|
| Long-run | wti$^+$ | 0.718*** | 1.205*** | 0.165*** | -0.202*** | 0.742*** | 0.685*** | 0.041 | 0.004 | 0.774*** |
|  | pbvr | 0.166*** | 0.263*** | 0.132*** | 0.095*** | 0.174*** | 0.145*** | 0.147*** | 0.093*** | 0.190*** |
|  | $\rho_i$ | -0.597*** | -0.425*** | -0.826*** | -0.924*** | -0.570*** | -0.529*** | -0.760*** | -0.779*** | -0.699*** |
|  | wti$^+$ | -0.146 | -0.468** | -0.185 | -0.418** | -0.066 | -0.385* | -0.239 | -0.084 | -0.148 |
|  | pbvr | 33.39 | 15.61 | 93.68 | 50.03 | 24.20 | 51.60 | 89.35 | 77.64 | 16.70 |
|  | nim |  | 0.442 |  |  |  |  |  |  |  |
|  | nocf |  |  | -5.467 |  |  |  |  |  |  |
| Short-run | lr |  |  |  | 39.32*** |  |  |  |  |  |
|  | ta |  |  |  |  | -6.190 |  |  |  |  |
|  | gdp |  |  |  |  |  | 0.470 |  |  |  |
|  | bc |  |  |  |  |  |  | 0.430 |  |  |
|  | rq |  |  |  |  |  |  |  | 3.971 |  |
|  | cpi |  |  |  |  |  |  |  |  | 1.488 |
|  | c | 2.636 | -10.58 | -1.075 | 12.15 | 4.843 | 0.415 | -0.030 | 8.077 | 3.924 |
| Log Likelihood |  | -460.2 | -443.3 | -446.0 | -428.7 | -433.1 | -434.0 | -436.1 | -422.7 | -400.8 |

*Note: \*\*\*, \*\*, \* indicates significance at 1%, 5% and 10%.*

Table A3. Results for the oil price negative shocks – robustness (z2)

|  | z2 | Model 1 | Model 2 | Model 3 | Model 4 | Model 5 | Model 6 | Model 7 | Model 8 | Model 9 |
|---|---|---|---|---|---|---|---|---|---|---|
| Long-run | wti- | -0.026 | -0.064*** | -0.309*** | -0.082*** | 0.432*** | -440.8*** | 0.014*** | -0.028* | -0.008 |
|  | pbvr | 0.154*** | 0.164*** | 0.100*** | 0.132*** | 0.202*** | 0.030*** | 0.141*** | 0.075*** | 0.204 |
|  | $\rho_i$ | -0.728*** | -0.662*** | -0.855*** | -0.955*** | -0.488* | -0.591*** | -0.708*** | -0.665*** | -0.467* |
|  | wti- | 0.077 | -0.312 | -0.059 | 0.012 | -0.017 | 0.510 | 0.081 | 0.082 | 0.141 |
|  | pbvr | 77.56 | 6.740 | 32.19 | -16.07 | -37.16 | -3.759 | 129.8 | 66.46 | 27.61 |
|  | nim |  | 0.213 |  |  |  |  |  |  |  |
|  | nocf |  |  | -0.761 |  |  |  |  |  |  |
| Short-run | lr |  |  |  | 24.21** |  |  |  |  |  |
|  | ta |  |  |  |  | 1.649 |  |  |  |  |
|  | gdp |  |  |  |  |  | 0.997* |  |  |  |
|  | bc |  |  |  |  |  |  |  | 0.271 |  |
|  | rq |  |  |  |  |  |  |  | -3.866 |  |
|  | cpi |  |  |  |  |  |  |  |  | 22.71 |
|  | c | 6.352 | 0.827 | 7.436 | 11.84 | 7.440 | 3.745 | 6.328 | 6.994 | 9.364 |
| Log Likelihood |  | -460.3 | -432.5 | -424.3 | -428.7 | -416.3 | -440.8 | -427.6 | -425.5 | -419.2 |

Note: ***, **, * indicates significance at 1%, 5% and 10%.

Table A4. Results for the oil price positive shocks – robustness (re-sampling)

|  | z1 | Model 1 | Model 2 | Model 3 | Model 4 | Model 5 | Model 6 | Model 7 | Model 8 | Model 9 |
|---|---|---|---|---|---|---|---|---|---|---|
| Long-run | wti+ | 0.668*** | 0.699*** | 0.776*** | 0.579*** | 0.181** | 0.150*** | -0.537*** | -0.144 | 2.230*** |
|  | pbvr | 0.123*** | 0.127*** | 0.104*** | 0.179*** | 0.418*** | 0.113*** | 0.092*** | 0.076*** | 0.100*** |
|  | $\rho_i$ | -0.643*** | -0.570*** | -0.607*** | -0.675*** | -0.649*** | -0.675*** | -0.513*** | -0.818*** | -0.592*** |
|  | wti+ | -0.182* | -0.439** | -0.240 | -0.451* | -0.057 | -0.271 | -0.355 | -0.054 | -0.301** |
|  | pbvr | -37.05 | 28.63 | 2.096 | -52.55 | 84.56 | 15.48 | 95.64 | 24.54 | 110.0 |
|  | nim |  | 0.560 |  |  |  |  |  |  |  |
|  | nocf |  |  | -5.699 |  |  |  |  |  |  |
| Short-run | lr |  |  |  | 29.77* |  |  |  |  |  |
|  | ta |  |  |  |  | 11.57 |  |  |  |  |
|  | gdp |  |  |  |  |  | 0.267 |  |  |  |
|  | bc |  |  |  |  |  |  | -1.071* |  |  |
|  | rq |  |  |  |  |  |  |  | 8.164** |  |
|  | cpi |  |  |  |  |  |  |  |  | -3.583* |
|  | c | 2.603 | -6.706 | -5.952 | 4.082 | 7.372 | 3.079 | -5.549 | 8.742 | 1.808* |
| Log Likelihood |  | -436.3 | -426.3 | -423.2 | -404.7 | -413.6 | -4.13.5 | -418.3 | -393.5 | -403.5 |

Note: ***, **, * indicates significance at 1%, 5% and 10%.

Table A5. Results for the oil price negative shocks – robustness (re-sampling)

|  | z1 | Model 1 | Model 2 | Model 3 | Model 4 | Model 5 | Model 6 | Model 7 | Model 8 | Model 9 |
|---|---|---|---|---|---|---|---|---|---|---|
| Long-run | wti- | 0.024 | -0.131*** | 0.031 | -0.061* | -0.227*** | -0.155*** | 0.519*** | -0.061** | -0.028*** |
|  | pbvr | 0.081*** | 0.015*** | 0.077*** | 0.732*** | 0.009 | 0.035*** | 0.079*** | 0.050*** | 0.066*** |
|  | $\rho_i$ | -0.710*** | -0.537*** | -0.638*** | -0.758*** | -0.389** | -0.555*** | -0.671*** | -0.700*** | -0.751*** |
|  | wti- | 0.128 | -0.135 | 0.116 | -0.062 | 0.227** | 0.599* | -0.022 | 0.143 | 0.131 |
|  | pbvr | 4.764 | -59.68 | 13.53 | -22.35 | -17.93 | -77.70 | 49.08 | 6.558 | 15.33 |
|  | nim |  | 0.370 |  |  |  |  |  |  |  |
|  | nocf |  |  | -5.637 |  |  |  |  |  |  |
| Short-run | lr |  |  |  | 22.49 |  |  |  |  |  |
|  | ta |  |  |  |  | 0.151 |  |  |  |  |
|  | gdp |  |  |  |  |  | 1.111* |  |  |  |
|  | bc |  |  |  |  |  |  | 0.057 |  |  |
|  | rq |  |  |  |  |  |  |  | 7.656* |  |
|  | cpi |  |  |  |  |  |  |  |  | 8.462 |
|  | c | 6.276 | 3.020 | -0.110 | 17.93* | 6.255 | 1.310 | 11.48* | 7.267 | 7.464 |
| Log Likelihood |  | -444.2 | -408.1 | -414.7 | -411.2 | -415.1 | -413.8 | -422.8 | -415.0 | -397.5 |

Note: ***, **, * indicates significance at 1%, 5% and 10%.